\documentclass[12pt]{article}

\usepackage{amsmath,amssymb,amsfonts,amsbsy}
\usepackage{cite}
\usepackage{graphicx}
\usepackage{wrapfig}
\usepackage{epsfig}

\usepackage{bbm} 
\usepackage{bm} 
\usepackage{color}                                                       %
\usepackage{dsfont} 
\usepackage{latexsym} 
\usepackage{lscape} 
\usepackage{mathrsfs} 
\usepackage{morefloats} 
\usepackage{slashed} 
\usepackage{psfrag}

\DeclareFontFamily{OT1}{mygreek}{}%
\DeclareFontShape{OT1}{mygreek}{m}{n}{<->omsegr}{}%
\DeclareFontShape{OT1}{mygreek}{b}{n}{<->omsegrb}{}%
\DeclareFontShape{OT1}{mygreek}{m}{it}{<->omsegri}{}%
\DeclareFontShape{OT1}{mygreek}{bx}{n}{<->sub * mygreek/b/n}{}%
\DeclareFontShape{OT1}{mygreek}{m}{sl}{<->sub * mygreek/m/it}{}%
\DeclareSymbolFont{Greekrm}{OT1}{mygreek}{m}{n} 
\DeclareSymbolFont{Greekbf}{OT1}{mygreek}{b}{n} 
\DeclareSymbolFont{Greekit}{OT1}{mygreek}{m}{it} 
\DeclareMathSymbol{\omegab}{\mathalpha}{Greekbf}{119}

\begin{document}
\addcontentsline{toc}{subsection}{{Title of the article}\\
{\it L.~Szymanowski}}

\setcounter{section}{0}
\setcounter{subsection}{0}
\setcounter{equation}{0}
\setcounter{figure}{0}
\setcounter{footnote}{0}
\setcounter{table}{0}

\begin{center}
\textbf{NLO CORRECTIONS TO TIMELIKE AND SPACELIKE DVCS}

\vspace{5mm}

B.~Pire$^{\,1}$, \underline{L.~Szymanowski}$^{\,2\,\dag}$ and
J. Wagner$^{\,2}$

\vspace{5mm}

\begin{small}
  (1) \emph{CPhT, \'Ecole Polytechnique,
CNRS, 91128 Palaiseau,     France} \\
  (2) \emph{National Center for Nuclear Research, Warsaw, Poland} \\
  $\dag$ \emph{E-mail: Lech.Szymanowski@fuw.edu.pl}
\end{small}
\end{center}

\vspace{0.0mm} 

\begin{abstract}
  Generalized Parton Distributions (GPDs) offer a new way to access the quark and gluon nucleon structure. We  advocate the need to supplement the experimental study of deeply virtual Compton scattering by its crossed version, timelike Compton scattering. We review recent progress in this domain, emphasizing the need to include NLO corrections to any phenomenological program to extract GPDs from experimental data. 
\end{abstract}

\vspace{7.2mm} 

The study of the internal structure of the nucleon has been the subject of many developments in the past decades and the concept of generalized parton distributions has allowed a breakthrough in the 3 dimensional description  of the quark and gluon content of hadrons. Hard exclusive reactions have been demonstrated to allow to probe the 
quark and gluon content of protons and heavier nuclei.

In this short review, we concentrate on  the complementarity of timelike and spacelike studies of hard exclusive processes, taking as an example the case of timelike Compton scattering (TCS) \cite{SzymanowskiBDP} where data at medium energy should be available at JLab@12 GeV and COMPASS, supplemented by higher energy data thanks both to the study of ultraperipheral collisions at RHIC and the LHC \cite{SzymanowskiPSW} and to a forthcoming electron-ion collider \cite{SzymanowskiarXiv:1108.1713}.

A considerable amount of theoretical and experimental work has 
been devoted to the study of deeply virtual Compton scattering (DVCS),
 i.e., $\gamma^* p \to \gamma p$, 
an exclusive reaction where generalized parton
distributions (GPDs) factorize from perturbatively calculable coefficient functions, when
the virtuality of the incoming photon is large enough.
 An extended research program for DVCS at JLab@12 GeV and 
Compass is now proposed to go beyond this first set of analysis. 
This will involve taking into account  next to leading order in $\alpha_s$ and next to 
leading twist contributions. We  advocate that it should be supplemented by the experimental study of  its crossed version, TCS, or even double DVCS \cite{SzymanowskiDDVCS} where both photons are off-shell.
 
\begin{wrapfigure}[10]{R}{50mm}
  \centering 
  \vspace*{-8mm} 
  \includegraphics[width=50mm]{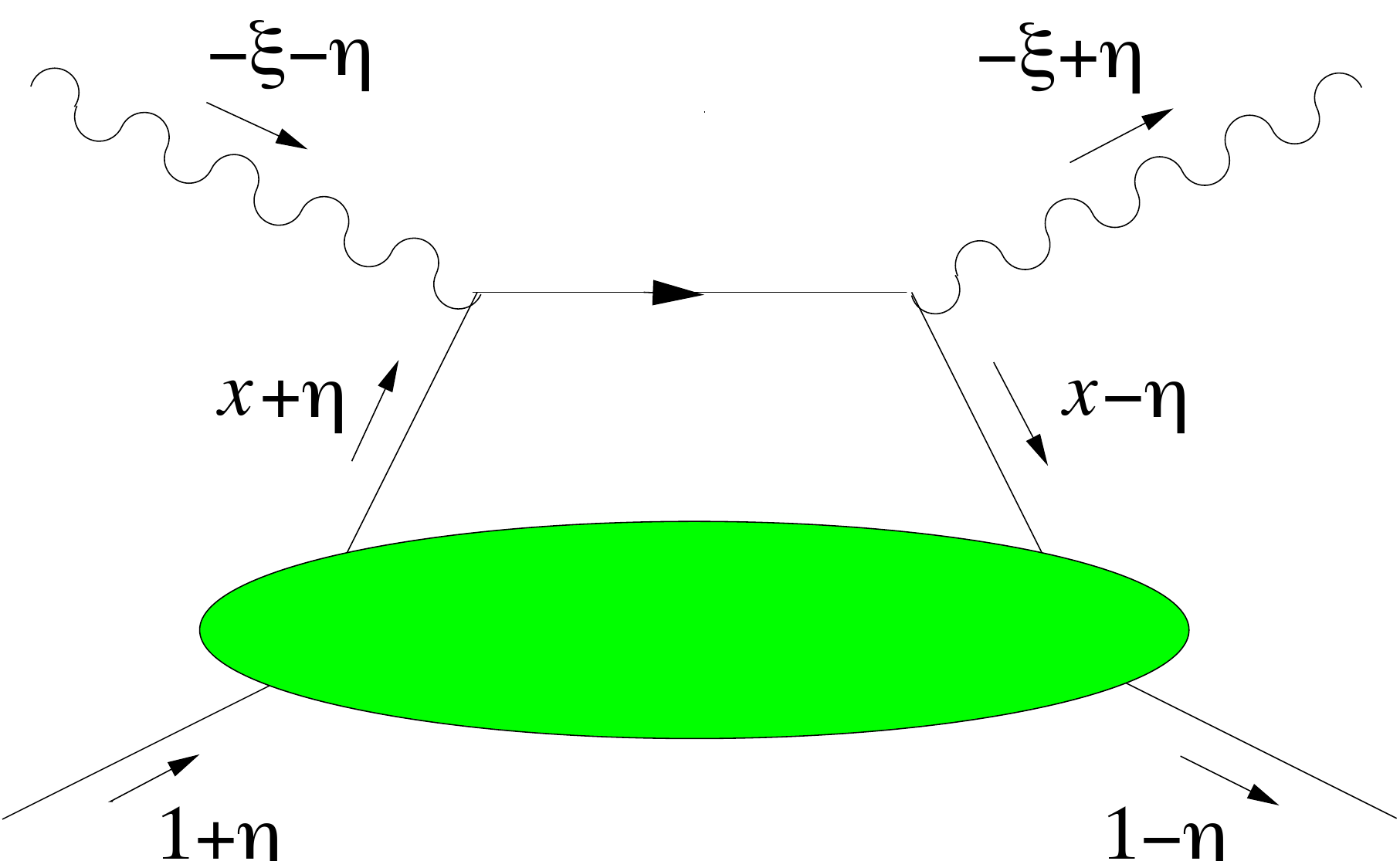}
  \caption{\footnotesize
The handbag mechanism controls both DVCS (where $\xi=\eta$) and TCS (where $\xi=-\eta$).}
  \label{SzymanowskiHandbag}
\end{wrapfigure}

 The physical process where to observe the inverse reaction, TCS \cite{SzymanowskiBDP},
 \begin{equation}
 \gamma(q) N(p) \to \gamma^*(q') N(p')
 \end{equation}
 is   the exclusive photoproduction of a
heavy lepton pair, $\gamma N \to \mu^+\!\mu^-\, N$ or $\gamma N \to
e^+\!e^-\, N$, at small $t = (p'-p)^2$ and large \emph{timelike} final state lepton pair squared mass $q'^2 = Q'^2$; TCS 
shares many features with DVCS. The generalized Bjorken variable in that case is $\tau = Q'^2/s $
 with $s=(p+q)^2$. One also defines $\Delta = p' -p$  ($t= \Delta^2$) and the skewness variables 
  $\eta$ and $\xi$\,, as
  $$\xi =  \frac{(q+q')^2}{2(p+p')\cdot (q+q')} ~~;~ \eta = - \frac{(q-q')\cdot (q+q')}{(p+p')\cdot (q+q')} 
  .$$
 For DVCS, $ \eta = \xi $ while for TCS, $\eta =- \xi \approx\,   \frac{ Q'^2}{2s  - Q'^2} $. 
At the Born order, both DVCS and  TCS amplitudes are described by the handbag diagram of Fig. 1 and its crossed version. They both interfere with a Bethe-Heitler QED process where the hadron structure enters through the well known nucleon form factors $F_1(t)$ and $F_2(t)$. The interference signal is a precise way to get an access to the DVCS and TCS amplitudes.

\begin{figure}[t]
\begin{center}
\includegraphics[width=0.4\textwidth]{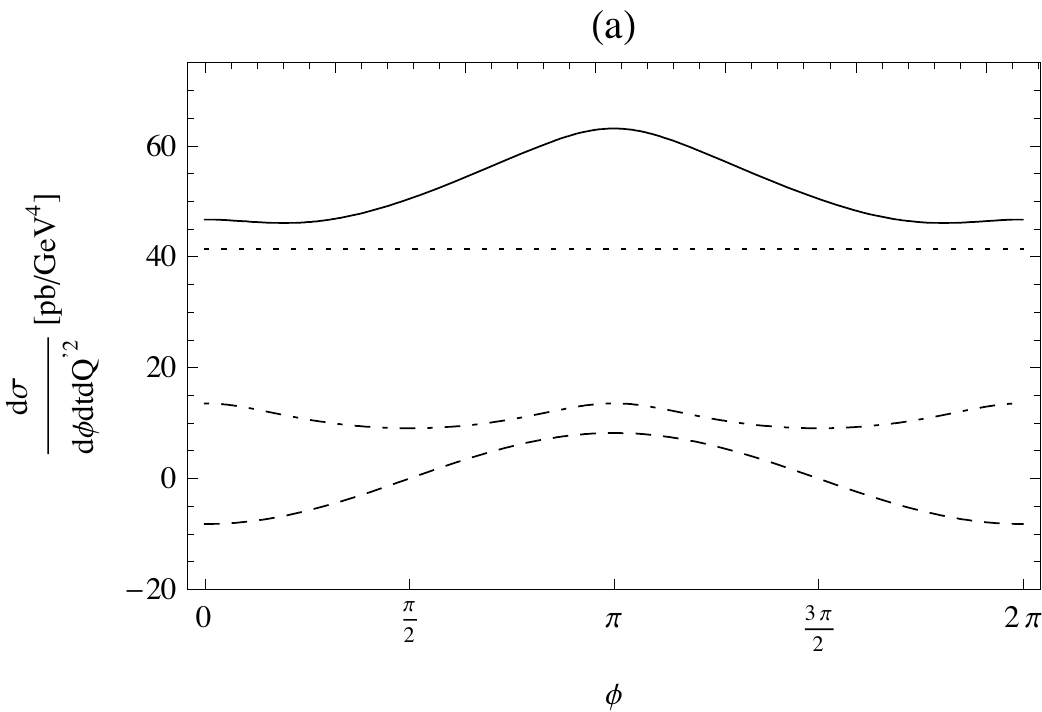}
\hspace{0.05\textwidth}
\includegraphics[width=0.4\textwidth]{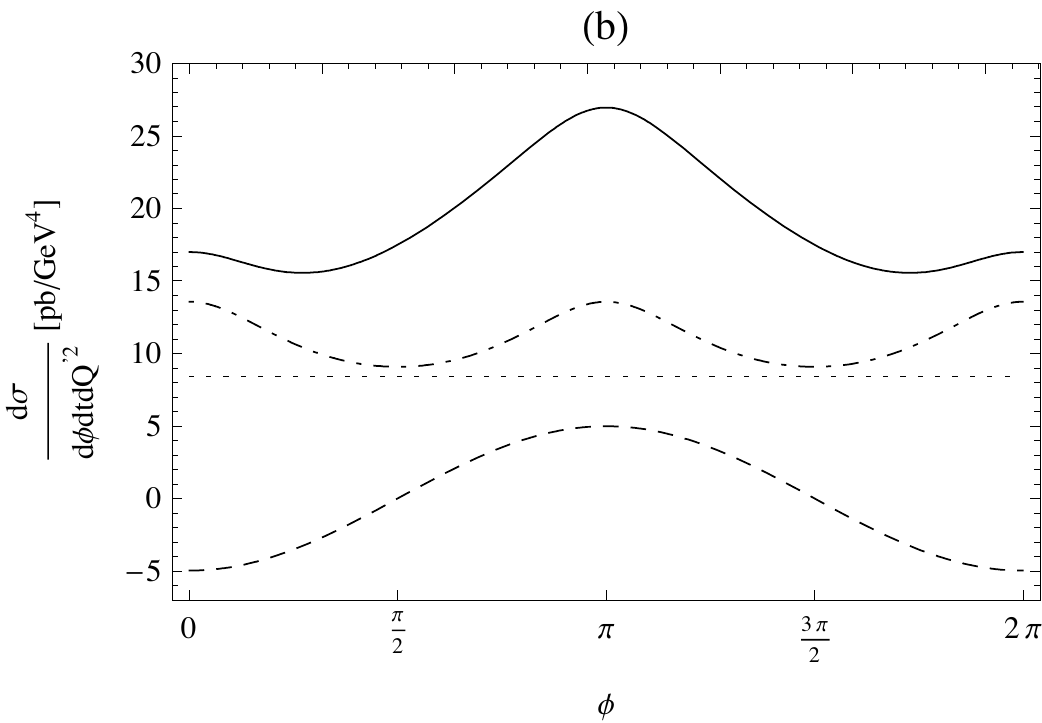}
\caption{
The differential cross sections (solid lines) for $t =-0.2 \,{\rm GeV}^2$, ${Q'}^2 =5 \,{\rm GeV}^2$ and integrated 
over $\theta = [\pi/4,3\pi/4]$, as a function of $\varphi$, for $s=10^7 \,{\rm GeV}^2$ (a), 
$s=10^5 \,{\rm GeV}^2$(b), with $\mu_F^2 = 5 \,{\rm GeV}^2$. We also display  the
Compton (dotted), Bethe-Heitler (dash-dotted) and Interference (dashed) contributions. 
}
\label{SzymanowskiInterf}
\end{center}
\end{figure}

The cross section for photoproduction in hadron collisions is given by:
\begin{equation}
\sigma_{pp}= 2 \int \frac{dn(k)}{dk} \sigma_{\gamma p}(k)dk \,,
\end{equation}
where $\sigma_{\gamma p} (k)$ is the cross section for the 
$\gamma p \to pl^+l^-$ process and $k$ is the photon energy. 
$\frac{dn(k)}{dk}$ is an equivalent photon flux. 
The relationship between $\gamma p$  energy squared $s$ and k is given by $
s \approx 2\sqrt{s_{pp}}k \nonumber$, 
where $s_{pp}$ is the proton-proton  energy squared. Figure \ref{SzymanowskiInterf} shows the interference contribution to the cross section in comparison to the Bethe Heitler and Compton processes, for various values of $\gamma N$ c.m. energy squared $s = 10^7 \,{\rm GeV}^2 $ and $10^5 \,{\rm GeV}^2$. We restrict the phase space integral to  $\theta = [\pi/4,3\pi/4]$ in order to avoid the overdominance of the QED process at forward angles. We observe  \cite{SzymanowskiPSW} that for large energies the Compton process dominates in these kinematics, whereas for $s=10^5 \,{\rm GeV}^2$ all contributions are comparable. This lowest order estimate shows that indeed TCS can be measured in ultraperipheral collisions at hadron colliders. For instance, we anticipate a rate of the order of $10^5$ TCS events per year at LHC with its nominal luminosity. This is mainly due to the large sea quark GPDs at very small $x$. It also calls for NLO corrections where  gluon GPDs start to contribute. 

Our calculations \cite{SzymanowskiPSW2} of NLO corrections show  important differences between the coefficient functions describing the TCS case and  those describing DVCS.  One defines the quark and gluon coefficient functions as
 \begin{eqnarray}
T^q = C_0^q+ C_1^q +\frac{1}{2} \log(\frac{ |Q^2|}{2\mu_F^2}) C_{coll}^q ~~~~~; ~~~~~
T^g =  C_1^g +\frac{1}{2} \log(\frac{ |Q^2|}{2\mu_F^2}) C_{coll}^g\;,
\nonumber
\label{eq:NLOTCSDVCS}
\end{eqnarray}
where $ C_0^q$ is the Born order coefficient function, $C^q_{coll} $ and $C^g_{coll} $ are directly related to the evolution equation kernels and $\mu_F$ is the factorization scale..

\begin{figure*}
\center
\includegraphics[width=0.4\textwidth]{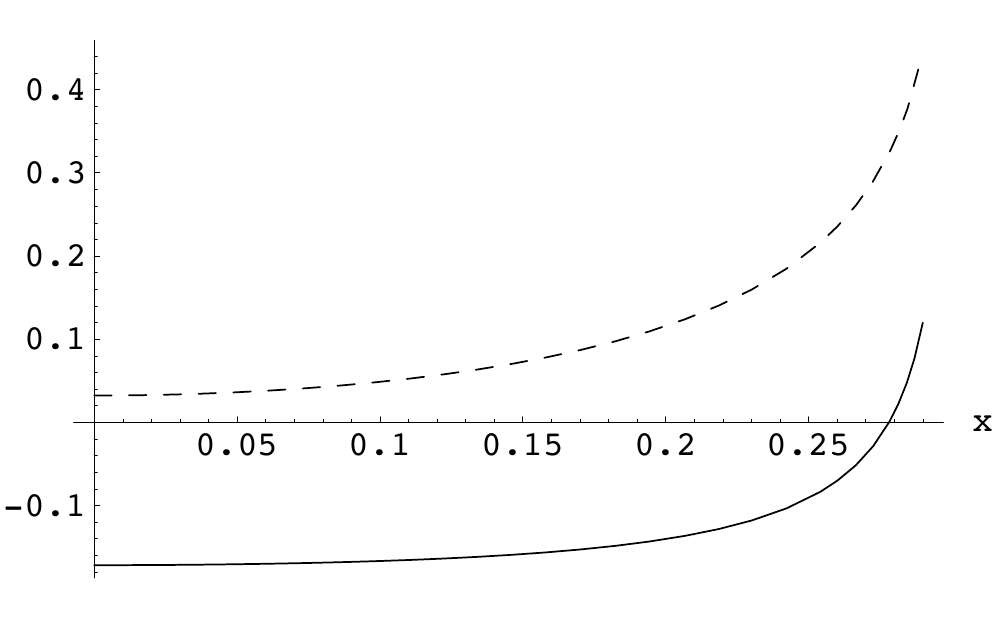} ~~~~~
\includegraphics[width=0.4\textwidth]{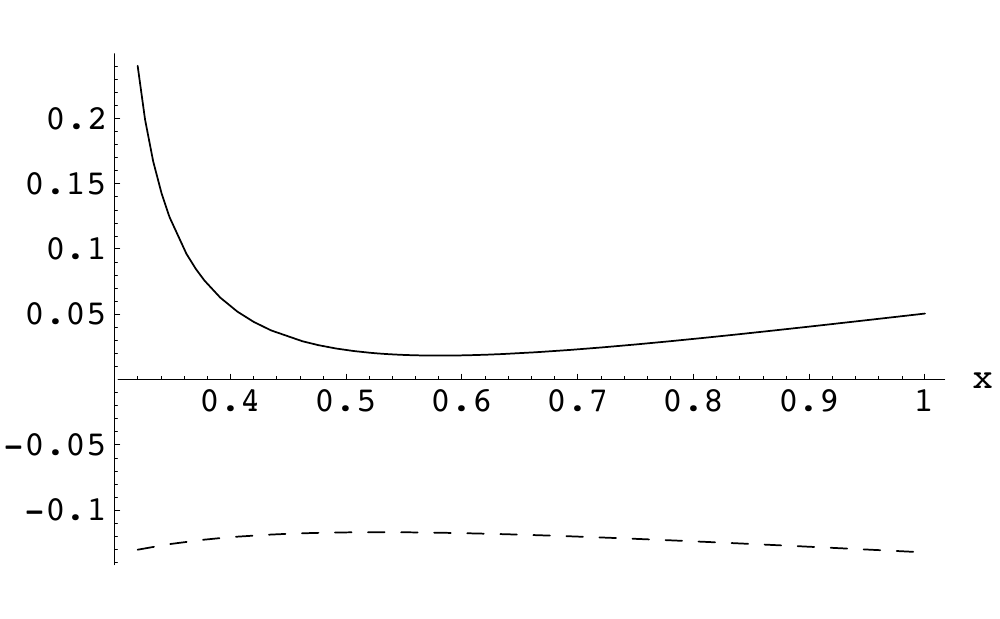} 
\includegraphics[width=0.4\textwidth]{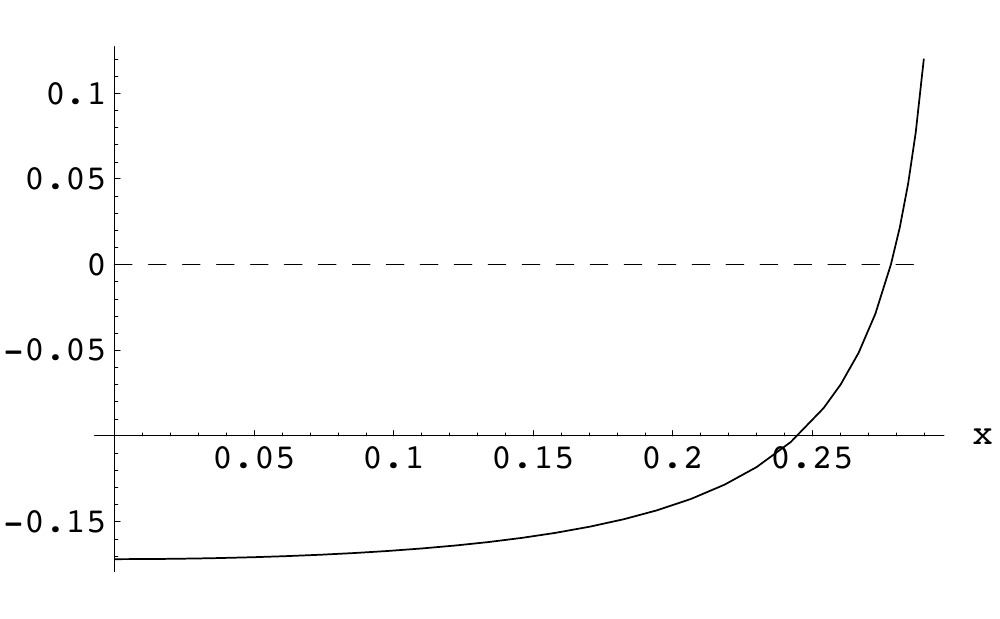} ~~~~
\includegraphics[width=0.4\textwidth]{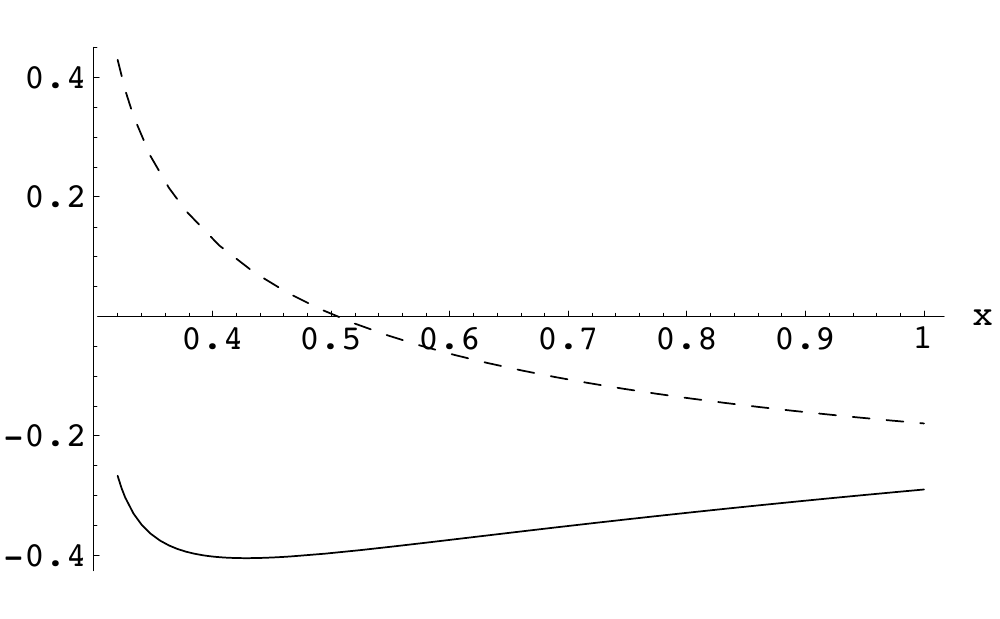} 
\caption{Real (solid line) and imaginary (dashed line) part of the ratio $R^q$ of the NLO quark coefficient function to the Born term in TCS (up) and DVCS(down) as a function of $x$ in the ERBL (left) and DGLAP (right) region for $\eta = 0.3$, for $\mu_F^2 = |Q^2|/2$.}
\label{SzymanowskiFig:ratio}
\end{figure*}

On Fig. \ref{SzymanowskiFig:ratio} we  show  the real and imaginary parts of the ratio $R^q = C_{1}^q/ C^q_{0}$ in timelike and spacelike Compton Scattering. We  fix $\alpha_s = 0.25$ and restrict the plots to the positive $x$ region, as the coefficient functions are antisymmetric in that variable. We see that in the TCS case, the imaginary part of the amplitude is  present in both the ERBL and DGLAP regions, contrarily to the DVCS case, where it exists  only in the DGLAP region. The magnitudes of these NLO coefficient functions are not small and neither is the  difference of the coefficient functions  ${C_{1(TCS)}^q}^* - C_{1(DVCS)}^q  $. The conclusion is that extracting the universal GPDs from both TCS and DVCS reactions requires much care. 

\begin{wrapfigure}[10]{R}{50mm}
  \centering 
  \vspace*{-8mm} 
  \includegraphics[width=50mm]{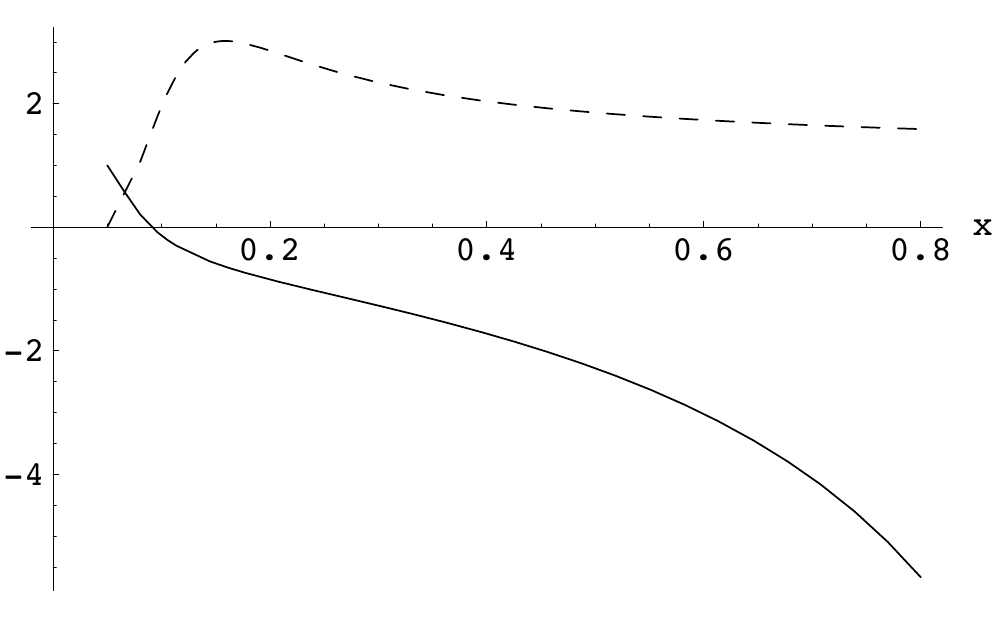}
  \caption{\footnotesize
Ratio of the real (solid line) and imaginary (dashed line) part of the NLO gluon coefficient function in TCS to the same quantity in DVCS  in the DGLAP  region .}
  \label{Szymanowskirg}
\end{wrapfigure}
An important part of NLO corrections to dVCS and TCS, especially at high energy, come from the gluon GPDs
Let us summurize our findings  on  the gluon coefficient functions. 
The  real parts of the gluon contribution are equal for DVCS and TCS in the ERBL region. 
The differences between TCS and DVCS emerges in the ERBL region through the imaginary part of the coefficient function which is non zero only for the TCS case and is of the order of the real part. 
In Fig. \ref{Szymanowskirg} we plot the ratio 
$\frac{C^g_{1(TCS)} }{ C^g_{1(DVCS)}}$
of the NLO gluon correction to the hard scattering amplitude in TCS to the one in  DVCS in the DGLAP  region for $\eta = 0.05$.

\begin{figure*}
\center
\includegraphics[width=0.4\textwidth]{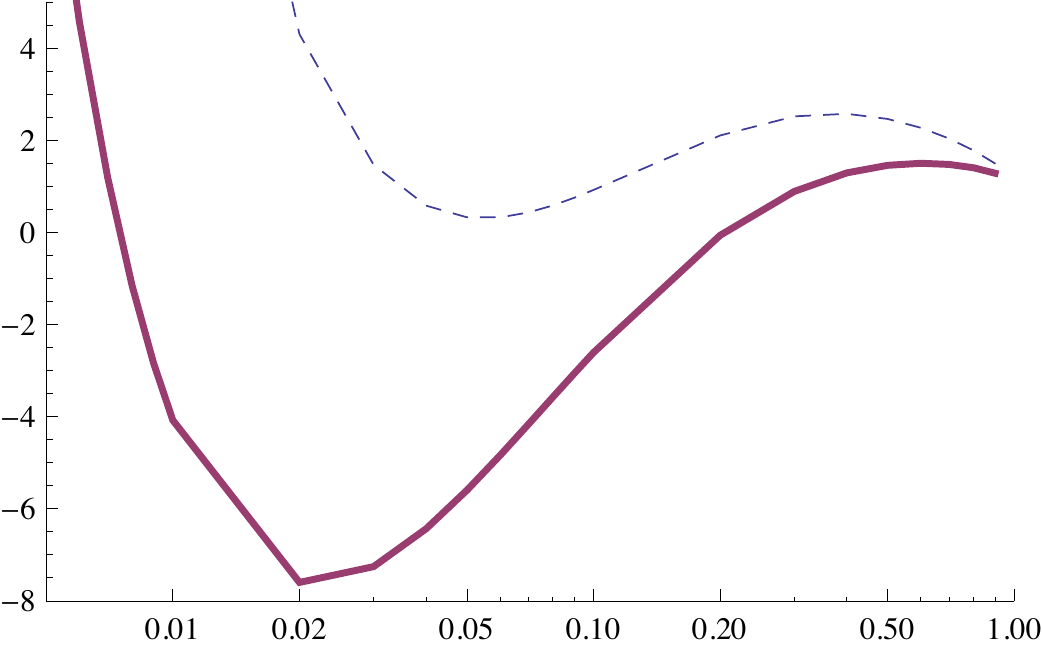} ~~~~~
\includegraphics[width=0.4\textwidth]{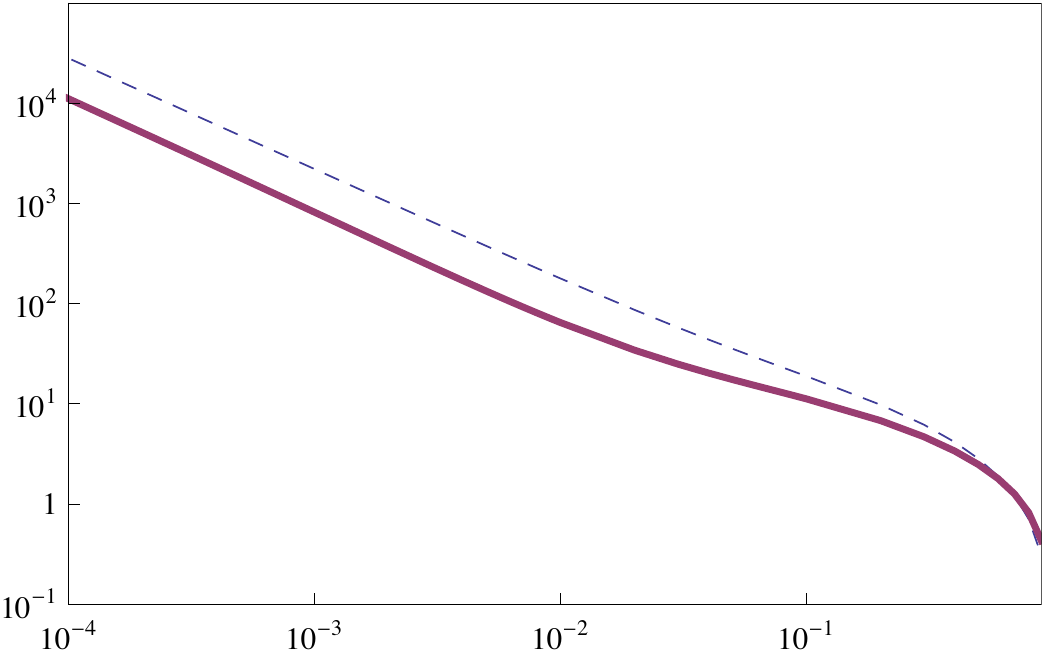} 
\caption{Real part (left) and imaginary part (right) of the DVCS Compton form factor ${\cal H}$ at LO (dashed) and  NLO (solid) as a function of $\xi$ for $\mu_F^2 = Q^2 = 4$ GeV$^2$.}
\label{SzymanowskiFig:CFF}
\end{figure*}

\vspace{.2cm}
To quantify the effects of this NLO calculations, we need to parameterize the quark and gluon GPDs and convolute them with the coefficient functions to get the Compton form factors that enter observable quantities, defined as
$${\cal H} = -\int_{-1}^1 dx \,[ \sum T^q(x,\xi,\eta)H^q(x,\xi,\eta) + T^g(x,\xi,\eta)H^g(x,\xi,\eta)],$$
fot the case of the GPD H, and corresponding definitions for other GPDs.
 Using the G-K model \cite{SzymanowskiGK}, we get, in the DVCS case, the results shown on Fig. \ref{SzymanowskiFig:CFF} which may be compared to earlier calculations \cite{Szymanowski9908337} . At small $\xi$, the imaginary part overdominates the real part, but the inclusion of NLO significantly diminishes its size. Note that the inclusion of NLO changes the sign of the real part in the valence region.
We are now calculating the corresponding Compton form factors for the timelike case.

\vspace{.2cm}
We acknowledge useful discussions with Franck Sabati\'e and Herv\'e Moutarde and the partial  support by the Polish Grant NCN No DEC-2011/01/D/ST2
/02069
and by the French-Polish Collaboration Agreement Polonium.

\end{document}